\documentclass[11pt, a4paper]{article}

\usepackage[utf8]{inputenc}     % For Unicode input encoding
\usepackage{amsmath}            % Advanced math typesetting
\usepackage{amssymb}            % Extra symbols
\usepackage{url}
\usepackage{geometry}           % To set page margins
\usepackage{setspace}           % For line spacing
\usepackage[natbibapa]{apacite}
\usepackage{natbib} % Load natbib for compatibility
\usepackage{graphicx}
\usepackage{listings}
\usepackage{fancyhdr}           % For header and footer
\usepackage{float}              % To handle figures and tables
\usepackage{hyperref} % Required for hyperlinks
\usepackage{booktabs} %toprule function
\usepackage[table]{xcolor} % For coloring table rows

\title{ouladFormat R Package: Preparing the Open University Learning Analytics Dataset for Analysis} % Article title
\author{
	Emma Howard \\
	\small Trinity College Dublin \\
	\small \texttt{emhoward@tcd.ie}
}
\date{}

\hypersetup{
	colorlinks   = true, %Colours links instead of ugly boxes
	urlcolor     = black, %Colour for external hyperlinks
	linkcolor    = black, %Colour of internal links
	citecolor   = black %Colour of citations
}

\usepackage{listings}

\definecolor{codedarkgray}{RGB}{105,105,105}
\definecolor{codeblack}{rgb}{0,0,0}
\definecolor{red}{rgb}{0.8,0.1,0.1}
\definecolor{backcolour}{rgb}{0.95,0.95,0.95}
\definecolor{codegreen}{rgb}{0,0.6,0}
\definecolor{codegray}{rgb}{0.5,0.5,0.5}
\definecolor{codepurple}{rgb}{0.58,0,0.82}

\lstdefinestyle{mystyle}{
	backgroundcolor=\color{backcolour},   
	commentstyle=\color{codegreen},
	keywordstyle=\color{codeblack},
	numberstyle=\tiny\color{codegray},
	stringstyle=\color{codeblack},
	basicstyle=\ttfamily\small,
	breakatwhitespace=false,         
	breaklines=true,    
	captionpos=b,                    
	keepspaces=true, 
	numbersep=5pt, showspaces=false,                
	showstringspaces=false, showtabs=false, tabsize=2}
\begin{document}

\flushbottom % Makes all text pages the same height

\maketitle % Print the title and abstract box
%\tableofcontents % Print the contents section

\thispagestyle{fancy} % Add header to first page

%\Keywords{OULAD, R package, learning analytics tools, educational data mining, MOOCs} 

\begin{abstract}
	
	Analysing educational data sets is fundamental to many fields of research focusing on improving student learning. However, large educational data sets are complex and can involve intensive preprocessing. These obstacles can be overcome through the development of educational tools which simplifies the preprocessing stages of analysis. The Open University Learning Analytics Dataset (OULAD), available online, contains data from 32,593 students across 22 module presentations at the Open University. This paper introduces the R software package ouladFormat; which loads and formats the OULAD for data analysis. The paper summarizes the ouladFormat R package and explains the different functions within the package. In addition, two case studies are provided which discuss how the OULAD and ouladFormat R package could be used when preparing for an educational study, and in the early identification of at-risk students. The package increases the accessibility of the OULAD for researchers, practitioners, and educators, and supports reproducibility and comparability of educational studies.
	
\end{abstract}

\section{Introduction}

Educational research is broadly defined as seeking `to describe, understand, and explain how learning takes place throughout a person's life and how formal and informal contexts of education affect all forms of learning' \citep{AERA}. A number of different areas of educational research have developed; each with their own specific goals, methodologies or population of interest. Those which focus on analysing, acquiring or harnessing insights from educational data sets include Academic Analytics, Educational Data Science, Educational Data Mining, and Learning Analytics (LA). For example, \citet[p. 1]{Motz} set forth the criteria of LA research to be research which uses data from learners engaged in education systems, which measures student learning, and improves the learning environments. Furthermore, \citet{Motz} analysed LA research from two main sources of publication, the Learning Analytics and Knowledge (LAK) conference and the Journal of Learning Analytics, to see if the studies were meeting these criteria. They found that 37.4\% of studies reviewed did not analyze data from learners in an education system.

To further the education field, it is essential that researchers and practitioners have access to real learner data. \citet{Mihaescu2021} identified 41 publicly available educational data sets. These include the Student Performance Dataset \citep{Cortez}, User Knowledge Modeling Data Set \citep{Kahraman}, Student Academics Performance Data Set \citep{Hussain} and the Open University Learning Analytics Dataset (OULAD) \citep{Kuz2017}. Studies which conduct the same statistical analysis on the these complex data sets may acquire different results owing to how the data are cleaned and preprocessed. \citet{crsra} highlight the complexity of massive open online course data and call for more tools to support researchers in analysing these data sets.  

\begin{figure}[ht!]
	\centering
	\includegraphics[width=140mm]{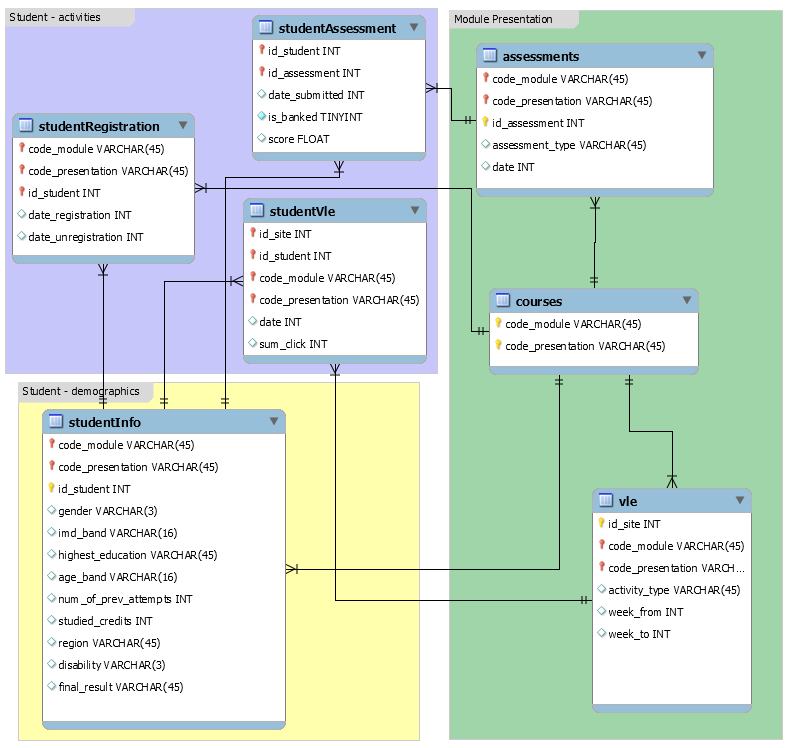}
	\caption{Figure from Kuzilek, Hlosta, and Zdrahal (n.d.) detailing the OULAD Database Schema} 
	\label{OULAD}
\end{figure}

The OULAD comes from the Open University, a world leader in providing distance learning. Based in the United Kingdom, in 2022/23, the Open University taught 150,619 students \citep{OU}. The OULAD \citep{Kuz2017} contains anonymized data from 22 presentations from 7 modules (courses). This includes demographic and registration information, assessment information and results, and Virtual Learning Environment (VLE) interactions represented by daily summaries of student clicks from 32,593 students. The OULAD is available online \citep{Kuz} across seven files under a database schema (see Figure~\ref{OULAD}). \citet{Kuz2017} provides definitions for each variable of the OULAD. To conduct statistical analysis, preprocessing is required. According to \citet{Mihaescu2021}, this intensive preprocessing is the main limitation of the OULAD. Hence, the ouladFormat R package \citep{Howard2024} was created to load, clean and format the OULAD for analysis as a single flat file (data set). Apart from streamlining the process of preparing the OULAD, the ouladFormat R package helps towards reproducibility and comparability of studies.

This paper explains how the ouladFormat R package can be used to prepare data for analysis. Section 2 summarizes the ouladFormat R package and explains the different functions within the package. In section 3, two case studies are provided where the ouladFormat is used to prepare data for subsequent data analysis. Section 4 concludes the paper.  

\section{The ouladFormat package}

\subsection{Overview}

The ouladFormat R package loads, cleans and formats the OULAD for data analysis. The overall aim of the package is to return the Open University data as a single flat data set/table/dataframe where each observation is a student. The package can be installed from CRAN using the command install.packages(``ouladFormat''). The package consists of ten functions (see Table~\ref{functions}). The functions have been grouped according to their purpose: 1) loading support, 2) loads specific data, 3) VLE format conversion, and 4) combines data sets (main function). The OULAD data can be considered under four data categories (demographics, assessment, registration, and VLE). The ouladFormat allows the user to load individual categories of data (see loads specific data functions) or for combination of data categories to be loaded (see combines data sets function) as a single flat file. The returned format of the data is a tibble \citep{Muller}, a type of dataframe. The main function, combined$\_$dataset(), draws on the other nine functions to return the formatted data for analysis. Hence, owing to this hierarchical structure, a user might only ever call upon the combined$\_$dataset() function. The other functions are useful for additionally flexibility. The ouladFormat R package does not provide functions for data analysis but the output can be combined with other packages for this purpose (see Section 3). 

While the OULAD includes data from 22 presentations across 7 modules \citep{Kuz2017}, this paper will predominately show results from the `2013J' presentation of the STEM module `DDD'. There were 1,503 students who have data available across the four data categories for the DDD-2013J presentation. Depending on the data categories requested by the user, the number of students per data set for a presentation will change. For example, for the DDD-2013J presentation, there are 1,938 students with registration data available. Student 3733 unregistered on day -8 (eight days before the start of the teaching semester), and therefore, there are no assessment data available for student 3733. When the data requested includes assessment data, student 3733 will not feature in the returned data set. There are 1,507 students with assessment data available. There are 1,507 students who have both assessment and registration data available.

\begin{table}[h!]
	\centering
	\small
 \caption{Description of Functions in the ouladFormat R Package}
  \label{functions}
  \begin{tabular}{lll} \hline
	    Function Name & Description edited from the R Package Help File & Function Purpose\\ \hline
	    \rowcolor{gray!30}  load$\_$github$\_$modified & \begin{tabular}[c]{@{}l@{}} Loads .RData files directly from a GitHub repository. \end{tabular} & Loading support \\ \hline
	    path$\_$to$\_$file & \begin{tabular}[c]{@{}l@{}} ouladFormat comes bundled with the sample student VLE data\\in its directory. This function makes it easy to access. \end{tabular}  &  Loading support\\ \hline
	   \rowcolor{gray!30} dataset$\_$demographics & \begin{tabular}[c]{@{}l@{}}Load and formats the studentInfo file from the OULAD\\for data analysis. \end{tabular} & Loads specific data \\ \hline
	     dataset$\_$registration & \begin{tabular}[c]{@{}l@{}}Load and formats the studentRegistration file from the\\ OULAD for data analysis. \end{tabular} & Loads specific data\\ \hline
	   \rowcolor{gray!30} dataset$\_$assessment & \begin{tabular}[c]{@{}l@{}} Load, combines and formats the assessment and \\studentAssessment files from the OULAD for data analysis.\end{tabular} & Loads specific data \\ \hline
	    dataset$\_$VLE$\_$time &\begin{tabular}[c]{@{}l@{}}Load and formats the studentVLE file from the OULAD\\for data analysis. \end{tabular}  & Loads specific data\\ \hline
	    \rowcolor{gray!30}dataset$\_$VLE$\_$activity & \begin{tabular}[c]{@{}l@{}} Load and formats the studentVLE and vle files from the\\OULAD for data analysis.\end{tabular} & Loads specific data\\ \hline
	     convert$\_$VLE & \begin{tabular}[c]{@{}l@{}} Converts the data format of VLE data from total view counts\\to binary, standardised or logarithmic view count data. \end{tabular} & \begin{tabular}[c]{@{}l@{}} VLE format\\conversion\end{tabular}  \\ \hline
	   \rowcolor{gray!30} VLE$\_$learning$\_$classification & \begin{tabular}[c]{@{}l@{}} Transform Open University VLE activities to classifications\\under either the FSLM, FSLSM, VARK or OLS mapping. \end{tabular} & \begin{tabular}[c]{@{}l@{}} VLE format\\conversion\end{tabular} \\ \hline
	    combined$\_$dataset & \begin{tabular}[c]{@{}l@{}} Main function. Combines multiple OULAD files into one\\tibble that is formatted for data analysis and where each row\\represents a unique student. \end{tabular} & Combines data sets \\ \hline
	\end{tabular}
	\end{table}

\subsection{Loading Support Functions}

The ouladFormat R package does not store the OULAD. Instead, the package features two functions (see Table~\ref{functions}) which help to load the OULAD into R. The user does not need to use these functions directly as they are called as part of the other package functions. For use of this package, the OULAD is stored as a series of .RData files on Github according to a database schema (see Figure 1). The load$\_$github$\_$modified(), based on the Rfssa function \citep{Haghbin}, which is based on a Stack Overflow post \citep{Hartman}, loads the .RData to a specified environment. The function was modified to allow the user to choose the environment in which the .RData files are loaded to. When a function, for example dataset$\_$demographics(), calls on the corresponding .RData, the data will be loaded into the local (function) environment rather than to the global environment. Functions return the requested subset of the OULAD as a formatted tibble. 

A subset of the OULAD studentVle data comes bundled with the package (5,000 rows from the AAA-2013J presentation). The path$\_$to$\_$file() function is adapted from the readxl$\_$example() function contained in the readxl R package \citep{readxl} and loads in the sample VLE data. The sample VLE data are used for VLE examples as manipulating the full VLE data takes longer than the recommended time for a CRAN package example. 

\subsection{Loading Specific Data Functions}

The five functions for loading specific data sets (see Table~\ref{functions}) are designed to have a similar structure through their sharing of three core parameters; module, presentation and repeat$\_$students. \citet{Kuz2017} specifies that the OULAD consists of three Social Sciences modules (AAA, BBB and GGG) and four STEM modules (CCC, DDD, EEE and FFF). The module parameter can specify an individual module or call upon all modules using the argument `All'. Each module has between two and four presentations present in the dataset. The presentations consist of 2013B, 2013J, 2014B, 2014J (B indicates a February start date and J indicates an October start date). Similarly, all presentations can be called upon using the argument `All'. The ouladFormat package will give an error if a selected presentation does not exist for a selected module. In the case of loading VLE data, there is no option to load `All' presentations of a module owing to the function design. This can be circumvented by loading each presentation individually and combining the outputs. The repeat$\_$students parameter indicates whether repeat students should be removed or kept. Each of the five loading specific data functions will be discussed in subsequent paragraphs. 

The dataset$\_$demographics() and dataset$\_$registration() functions, apart from returning the core arguments as outputs, return the OULAD studentInfo and studentRegistration data respectively. Minimal changes are made to the OULAD studentInfo and studentRegistration data. 

\begin{lstlisting}[language=R]
	# Example code for using the dataset_demographics() 
	# and dataset_registration() functions:
	
	# Return demographic data for students in module DDD and presentation 2013J:
	dataset_demographics(module = "DDD", presentation = "2013J", 
	repeat_students = "remove")
	# Return demographic data for students in module DDD (all presentations):
	dataset_demographics(module = "DDD", presentation = "All", 
	repeat_students = "remove")
	# Return registration data for students who undertook a module with a presentation 2013J:
	dataset_registration(module = "All", presentation = "2013J",
	repeat_students = "keep")
\end{lstlisting}

The dataset$\_$assessment() function, apart from having the three core parameters, has the parameters week$\_$begin, week$\_$end and na.rm. These parameters give the user greater flexibility over the output by allowing the user to specify the period in the semester that the returned assessment data relates to. For example, the user may wish to only include assessments in the returned tibbles that had due dates in the first 10 weeks of the module (see example code below). 

\begin{lstlisting}[language=R]
	# Example code and output for using the dataset_assessment() 
	# function: 
	
	> dataset_assessment(module = "DDD", presentation = "2013J", 
	+	repeat_students = "remove", week_begin = 1, week_end=10, 
	+	na.rm = FALSE)
	$assessment_data
	# A tibble: 2,379 X 11
	id_student code_module code_presentation id_assessment assessment_type    
	<chr>      <fct>       <fct>             <fct>         <fct>      
	1 102850      DDD         2013J             25349         TMA 
	2 102850      DDD         2013J             25348         TMA 
	3 103800      DDD         2013J             25348         TMA 
	4 103800      DDD         2013J             25349         TMA 
	5 104643      DDD         2013J             25349         TMA 
	6 104643      DDD         2013J             25348         TMA
	7 104772      DDD         2013J             25348         TMA
	8 104772      DDD         2013J             25349         TMA
	9 105523      DDD         2013J             25348         TMA   
	10 105523     DDD         2013J             25349         TMA   
	# 2,369 more rows
	# 6 more variables: date <dbl>, weight <dbl>, date_submitted <dbl>, 
	#	is_banked <fct>, score <dbl>, reactivity <dbl>
	# Use `print(n = ...)` to see more rows
	
	$assessments
	# A tibble: 7 x 7
	code_module code_p... id_a... assessment_type  date weight  week
	<chr>        <chr>   <int>   <chr>           <int>  <dbl> <dbl>
	1 DDD         2013J   25348   TMA                25   10       4
	2 DDD         2013J   25349   TMA                53   12.5     8
	3 DDD         2013J   25350   TMA                88   17.5    13
	4 DDD         2013J   25351   TMA               123   20      18
	5 DDD         2013J   25352   TMA               165   20      24
	6 DDD         2013J   25353   TMA               207   20      30
	7 DDD         2013J   25354   Exam              261  100      38
	
	$assessment_performance
	# A tibble: 1,271 X 4
	id_student `25349` `25348` average_CA_score
	<dbl>   <dbl>   <dbl>            <dbl>
	1     102850      97      68             84.1
	2     103800      56      63             59.1
	3     104643      81      77             79.2
	4     104772      56      67             60.9
	5     105523      69      81             74.3
	6    1080206      81      93             86.3
	7     108789      NA      27             12  
	8    1103608      87      84             85.7
	9     114558      81      81             81  
	10   1187586      92      84             88.4
	# 1,261 more rows
	# Use `print(n = ...)` to see more rows
	
	$assessment_reactivity
	# A tibble: 1,271 X 3
	id_student    `25349` `25348`
	<chr>       <dbl>   <dbl>
	1 102850           0       8
	2 103800           0       0
	3 104643           1       5
	4 104772           1       1
	5 105523           0       0
	6 1080206          6       5
	7 108789          NA      -2
	8 1103608          7       3
	9 114558           1       8
	10 1187586         2       6
	# 1,261 more rows
	# Use `print(n = ...)` to see more rows
	
	$module
	[1] "DDD"
	
	$presentation
	[1] "2013J"
	
	$repeat_students
	[1] "remove"
\end{lstlisting}

The dataset$\_$assessment() function returns a list of the three core parameters and the four tibbles: 1) assessment$\_$data, 2) assessments, 3) assessment$\_$performance, and 4) assessment$\_$reactivity. The assessment$\_$data, assessment$\_$performance, and assessment$\_$reactivity are all filtered based on the week$\_$begin and week$\_$end parameters. The assessment$\_$data is a tibble based on the combined OULAD files of studentAssessment and assessments (see Figure~\ref{OULAD}), and filtered based on the inputs. The assessments tibble, as seen above, returns the full list of assessments for the selected module-presentation combination. This tibble is particularly beneficial for the user in learning about the module as continuous assessment drives engagement. For DDD-2013J, there were six continuous assessments; all marked by a tutor. These were weighted 10\%, 12.5\%, 17.5\%, 20\%, 20\% and 20\%, and were due in weeks 4, 8, 13, 18, 24, and 30 respectively. The assessment$\_$performance tibble is a tibble where each row represents a unique student and their scores in the range of 0-100 for different assessment items (see id$\_$assessment in the assessments tibble). When a specific module presentation is called, for example DDD-2013J, the ouladFormat package calculates students' average weighted continuous assessment (CA) score based on the CA variables included in the filtered data and returns it as part of the assessment$\_$performance tibble. Any (end-of-semester) examination data is not included in the calculation. According to \citet{Kuz2017}, if a student does not submit the assessment, a NA is recorded for that assessment. For this calculation, the NAs may be excluded or replaced by 0 (default) through using the na.rm parameter. The fourth tibble, the assessment$\_$reactivity tibble, returns the reactivity for each assessment. \citet{Treuiller} define reactivity for the OULAD as the delay between the date the assessment is returned and the deadline (in days). Negative numbers indicate overdue assessments.

The OULAD studentVle file provides information on how many times a specific resource (coded as the variable id$\_$site) for a specific date was accessed by a student. Using the OULAD vle file, the classification of the resource can be identified (e.g., homepage, oucontent, and forumng). Formatted VLE data can be returned using the dataset$\_$VLE$\_$activity() and dataset$\_$VLE$\_$time() functions. dataset$\_$VLE$\_$time() returns tibbles of the number of times students accessed VLE resources per day and per week. dataset$\_$VLE$\_$activity() returns a tibble of the number of times students accessed VLE resources by activity classification. In addition to the core parameters, both these functions have the parameters week$\_$begin, week$\_$end and example$\_$data (see Section 2.2). Similar to the dataset$\_$assessment() function, week$\_$begin and week$\_$end allow the user to control what period, or weeks of the semester, the VLE data relates to in the returned tibbles. 

\begin{lstlisting}[language=R]
	# Example code for using the dataset_VLE_time() and 
	# dataset_VLE_activity() functions:
	
	# See Section 2.2 on example student VLE data 
	# (subset of AAA-2013J):
	dataset_VLE_time(example = TRUE)
	
	# Return all VLE data available for students in module DDD and presentation 2013J:
	dataset_VLE_time(module = "DDD", presentation = "2013J", 
	repeat_students = "keep", week_begin = -4, week_end = 39, 
	example_data = FALSE)
	
	# Return VLE data for students in module BBB and presentation 2013J for weeks 1-13:
	dataset_VLE_time(module = "BBB", presentation = "2013J",
	repeat_students = "remove", week_begin = 1, week_end = 13, 
	example_data = FALSE)
	
	# Return VLE activity data for students in DDD-2013J with weeks prior to module start excluded:
	dataset_VLE_activity(module = "DDD", presentation = "2013J", 
	repeat_students = "keep", week_begin = 1, week_end = 39, 
	example_data = FALSE)
\end{lstlisting}

Apart from returning tibbles, the dataset$\_$VLE$\_$activity() and dataset$\_$VLE$\_$time() return the core arguments as well as values for week$\_$begin and week$\_$end. However, the returned values for week$\_$begin and week$\_$end may differ from the inputted values. The inputted values represent the time period of VLE data that users request be returned in the outputted tibbles. The returned values represent the week$\_$begin and week$\_$end period that is in the returned tibbles. In the example below, the code requests VLE data for DDD-2013J for between week -6 (six weeks before the start date) and week 2 of the semester inclusive. However, students only started accessing the module VLE in week -3, and therefore the returned value for week$\_$begin is -3; this is evident in the tibbles outputted as well.

\begin{lstlisting}[language=R]
	# Example code and output for using the dataset_VLE_time() 
	# function: 
	
	> dataset_VLE_time(module = "DDD", presentation = "2013J", 
	+	repeat_students = "remove",  week_begin = -6, 
	+ week_end = 2, example_data = FALSE)$filtered_data
	# A tibble: 110,953 x 6
	id_student    code_module code_presentation id_site  date sum_click
	<chr>      <fct>       <fct>               <int>  <dbl>     <dbl>
	1  102850     DDD         2013J              674078    -2         2
	2  102850     DDD         2013J              674186    -2         1
	3  102850     DDD         2013J              673740    -2         5
	4  102850     DDD         2013J              674054    10         1
	5  102850     DDD         2013J              673981    -2         2
	6  102850     DDD         2013J              673537    -2         1
	7  102850     DDD         2013J              673724   -12         1
	8  102850     DDD         2013J              674075    -2         1
	9  102850     DDD         2013J              674158    -2         1
	10 102850     DDD         2013J              674306    -2         1
	#  110,943 more rows
	#  Use `print(n = ...)` to see more rows
	
	$daily_data
	# A tibble: 1,472 x 33
	id_student     `-18` `-17` `-16` `-15` `-14` `-13` `-12` `-11` 
	<chr>          <dbl> <dbl> <dbl> <dbl> <dbl> <dbl> <dbl> <dbl>  
	1  1103608      105     0    17    28     1    11     0     0  
	2  122739        11     0     0     0     0     1     0     0  
	3  123957        14     5     3     0     0     0     7     0  
	4  126394        47     0     0     0     0     0     0     0 
	5  130467         2     4    16     0     0     0     0     4  
	6  130671        10     7    10    10    34    26     7     1  
	7  130837        63     0     0     0     0     0     0     0 
	8  131222        20    26     0     0     5     1     0     0  
	9  1345808       16     0     0     0     0     0     0     0   
	10 1440467       66    30    50     0     0     0     0     0  
	# 1,462 more rows
	# 24 more variables: `-10` <dbl>, `-9` <dbl>, `-8` <dbl>,
	#   `-7` <dbl>,`-6` <dbl>, `-5` <dbl>, `-4` <dbl>, `-3` <dbl>,
	#   `-2` <dbl>, `-1` <dbl>, `0` <dbl>, `1` <dbl>, `2` <dbl>,
	#   `3` <dbl>,  `4` <dbl>, `5` <dbl>, `6` <dbl>, `7` <dbl>,
	#  `8` <dbl>, `9` <dbl>, `10` <dbl>, `11` <dbl>, `12` <dbl>,
	#   `13` <dbl>
	# Use `print(n = ...)` to see more rows
	
	$weekly_data
	# A tibble: 1,472 x 6
	id_student    `Week_pre-3` `Week_pre-2`   `Week_pre-1` Week1 Week2
	<chr>                <dbl>        <dbl>          <dbl> <dbl> <dbl>
	1  104643                5           70           32    61    36
	2  104772                7           66           35    61   186
	3  1103608             150           18          257    75    75
	4  114558               22           66           41   187    80
	5  118887                7            2            0     0     0
	6  122739               11            1           24    10    52
	7  123957               22           11           28    86    31
	8  124387               14            0            0    42     0
	9  126394               47            0            0     0     0
	10 126812               17          115          106   106     9
	# 1,462 more rows
	# Use `print(n = ...)` to see more rows
	
	$module
	[1] "DDD"
	
	$presentation
	[1] "2013J"
	
	$repeat_students
	[1] "remove"
	
	$week_begin
	[1] -3
	
	$week_end
	[1] 2
\end{lstlisting}

\subsection{VLE Format Conversion Functions}

The defaults of dataset$\_$VLE$\_$activity() and dataset$\_$VLE$\_$time() are to return tibbles with the number of times students accessed resources per day, per week or per activity type. However, a number of statistical methods (e.g., $k$-means) are sensitive to variances across variables, and other data formats may be needed. convert$\_$VLE() quickly converts the VLE data to either binary, standardised by variable, standardised globally, or logarithmic format. Some studies have mapped the OULAD activities to different classifications. For example, \citet{Balti} mapped the activity classifications to the VARK Learning Style Model \citep{Fleming}. The VLE$\_$learning$\_$classification() function maps the VLE activities according to different studies (see function's help file), and returns the mapping as well as a tibble of the VLE data according to the chosen mapping.

\begin{lstlisting}[language=R]
	# Example code for using the convert_VLE() and 
	# VLE_learning_classification functions:
	
	# Loads and standardises sample VLE data by variable
	VLE_data = dataset_VLE_activity(example_data = TRUE)$resource_data
	convert_VLE(VLE_data, conversion = "standardise1")
	
	# Loads sample VLE data and maps activities under the 
	# Felder-Silverman Learning Style Model:
	VLE_data = dataset_VLE_activity(example_data = TRUE)$resource_data
	VLE_learning_classification(VLE_data, classification = "VARK")
	
	# See more mappings
	?VLE_learning_classification
\end{lstlisting}

\subsection{Combines Data Sets (Main Function)}

The main function of the ouladFormat package is combined$\_$dataset(), and this function calls on the other nine functions in the package to allow the user to easily return their required data set as a single tibble. In addition to the parameters previously seen, the demographics, registration, assessment and VLE parameters allow the user to specify if each of the corresponding data categories should be included in the returned tibble, and in the case of the VLE parameter further flexibility is provided; with even more flexibility possible through the VLE$\_$clicks parameter. Additionally, there is a withdrawn$\_$students parameter to indicate whether withdrawn students should be removed. When set to `remove', this does not necessarily remove all students who withdrew from the module, rather it removes students who withdrew from the module up to and including the week indicated by the week$\_$end argument (see Section 3.2 for example). Note for the combined$\_$dataset(), repeat students are always removed when presentation set to `All'.

\begin{lstlisting}[language=R]
	# Usage of combined_dataset():
	combined_dataset(
	module = c("AAA", "BBB", "CCC", "DDD", "EEE", "FFF", "GGG"),
	presentation = c("2013J", "2014J", "2013B", "2014B", "All"),
	repeat_students = c("remove", "keep"),
	withdrawn_students = c("remove", "keep"),
	demographics = FALSE,
	registration = FALSE,
	VLE = c("omit", "daily", "weekly", "activity", "FSLM", "FSLSM",
	"OLS", "VARK"),
	VLE_clicks = c("total", "binary", "standardise1", "standardise2", 
	"logarithmic"),
	week_begin = -4,
	week_end = 39,
	assessment = FALSE,
	na.rm = FALSE,
	example_data = FALSE
	)
\end{lstlisting}

\section{Case Studies}

\subsection{Case Study 1: Preparation for an educational study}

\begin{figure}[ht!]
	\centering
	\includegraphics[width=120mm]{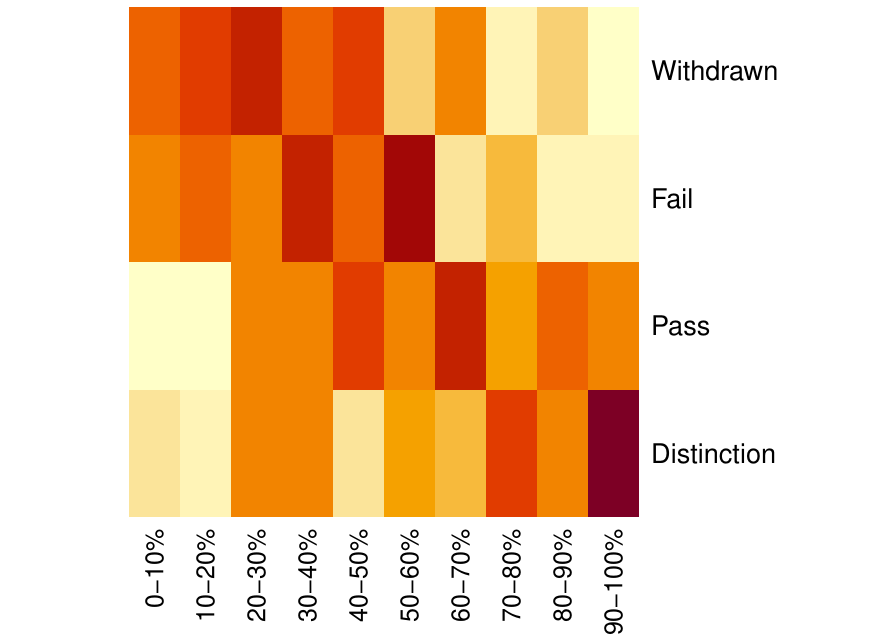}
	\caption{Heatmap illustrating for each final result category, the proportion of students within each imd band. Darker colours indicate a higher proportion of students.} 
	\label{case1}
\end{figure}

Researchers, or institution staff, may wish to conduct an educational study. Pre-registration of data analysis plans are becoming more common, and any data to be collected should be determined in advance of conducting a study. In preparation, researchers, through exploring the OULAD, can improve their data analytics skills, learn about relationships underpinning educational data to inform their data collection and knowledge, and gather evidence for their own educational hypotheses for ethics or funding applications. For example, a researcher may wish to examine whether extra supports for students from low-income backgrounds would decrease student drop out and failure rates. To acquire funding for this project, researchers may need to show evidence that there exists a difference in achievement levels between students from low-income and high-income backgrounds. One source of evidence could be showing the relationship between the OULAD final$\_$result and imd$\_$band variables. The imd$\_$band variable gives the index of multiple deprivation (imd) band of the place where the student lived during a module-presentation \citep{Kuz2017}. Figure~\ref{case1} shows that of those students who withdrew or failed, a higher proportion came from lower imd bands. The significance of the relationship between the final result and the imd band is confirmed by a $\chi^2$ statistic of 161.74 with 27 degrees of freedom. 

\begin{lstlisting}[language=R]
	# Load ouladFormat
	require(ouladFormat) 
	
	# Load the demographic data 
	case1 <- dataset_demographics(module = 'DDD', presentation = 'All', 
	repeat_students = 'keep')$studentInfo
	
	# Format final result variable 
	# (makes Figure 2 nicer as categories ordered)
	case1$final_result = factor(case1$final_result, 
	levels = c("Distinction", "Pass", "Fail", "Withdrawn"))
	
	# Examine the relationship between final result and the index of 
	# multiple deprivation band of the place where the student lived 
	# during the presentation
	cross_tab = table(case1$final_result, case1$imd_band) 
	cross_tab 
	chisq.test(cross_tab) # Run X^2 test
	# Creates Figure 2
	heatmap(cross_tab, Rowv = NA, Colv = NA, scale = "row", cexRow = 1.3) 
\end{lstlisting}

\subsection{Case Study 2: Early identification of at-risk students}

An early warning system in education involves identifying students at risk of failing or dropping out of a module/course and providing them with an intervention to help support them in their studies. Ideally, at-risk students would be identified as early as possible in the teaching semester in order to provide effective interventions \citep{Howard}. At-risk students may be identified by academic staff or through statistical methods such as prediction modelling. While many studies have applied prediction modelling to the OULAD for various motivations \citep{Jin}, \citet{Drousiotis} aimed to identify at-risk students by predicting students' final results (withdrawn, fail, pass and distinction) as early as possible in the teaching semester. To achieve this, they used the explanatory variables of first assessment mark, highest educational level achieved, number of VLE clicks until the module starts, registration date, age, disability, gender, and number of previous attempts. Then they applied a 70\%/30\% split for training/test data and used the prediction algorithms of Decision Tree classifier, Random Forest, and Bayesian Additive Regression Trees. The code below demonstrates the approach taken by \citet{Drousiotis} for the DDD-2013J presentation using the Decision Tree Classifier and Random Forest algorithms. Overall, the Decision Tree classifier achieves the higher accuracy (0.5262 versus 0.5011). If the intention was to provide at-risk students with interventions, the Decision Tree classifier could be used to identify at-risk students (those predicted to fail or withdraw) in the next iteration of the module and subsequently interventions provided. 

Researchers may be interested in finding the best prediction model for identifying at-risk students and use the OULAD for this. The ouladFormat R package can aid comparisons between models as it makes it easier for studies to use the same data from the OULAD.   

\begin{lstlisting}[language=R]
	# Load required packages for case study
	require(ouladFormat)
	require(tidyverse)
	require(randomForest)
	require(caret)
	require(party)
	
	# Load in the DDD-2013J data, match variables used by 
	# Drousiotis, Shi and Maskell (2021) and remove 
	# redundant variables:
	case2 <- combined_dataset(module = 'DDD', presentation = '2013J', 
	withdrawn_students = "remove", repeat_students = 'keep', 
	demographics = TRUE,  assessment= TRUE, registration = TRUE, 
	VLE = "weekly",  VLE_clicks = 'total',
	week_begin = -4, week_end = 4)$dataset_combined %>% 
	mutate(clicks = `Week_pre-3`+`Week_pre-2`+`Week_pre-1`) %>% 
	rename(CA1 = `25348`) %>% # Rename first assessment variable
	select(!c(id_student:code_presentation, date_unregistration, 
	region, imd_band, studied_credits, average_CA_score:Week4)) 
	
	# Remove incomplete student records
	case2 <- case2[complete.cases(case2),]
	
	# Set seed to ensure reproducibility of results
	set.seed(123)
	
	# Implement 70%/30% data split
	trainIndex <- createDataPartition(case2$final_result, 
	p = 0.7, list = FALSE)
	trainData <- case2[trainIndex, ] %>% as.data.frame()
	testData <- case2[-trainIndex, ] %>% as.data.frame()
	
	# Implement Random Forest approach
	Model1 <- randomForest(final_result ~ ., data = trainData, 
	ntree=50)
	predictions1 <- predict(Model1, newdata = testData)
	confMatrix1 <- confusionMatrix(predictions1, testData$final_result)
	print(confMatrix1)
	
	# Implement Decision Tree approach
	Model2 <- ctree(final_result ~ ., data = trainData)
	predictions2 <- predict(Model2, newdata = testData)
	confMatrix2 <- confusionMatrix(predictions2, testData$final_result)
	print(confMatrix2)
\end{lstlisting}

\section{Conclusion}

Educational research benefits by having large real-life data sets, such as the OULAD \citep{Kuz2017}, freely available. However, large educational data sets are complex and can be a challenge for those working with them \citep{crsra}. Similar to the crsra R package \citep{crsra}, which supports researchers in cleaning and analysing Coursera's research data exports, the ouladFormat R package is a flexible tool to help researchers load and format the OULAD for analysis. The main function, combined$\_$dataset(), allows the user to easily return a subset of the OULAD as a single data set. The parameters of combined$\_$dataset() are designed to give the user flexibility. 

While the two case studies presented in Section 3 focused on examples of researchers or practitioners making use of the OULAD, the OULAD may also be of interest to educators teaching statistical analysis and methods. For example, a subset of the OULAD could be used as a data set for a multivariate analysis or machine learning module, or for a final year undergraduate project. For educators, we hope that the ouladFormat R package makes the OULAD accessible for teaching. For researchers, we hope the ouladFormat R package helps support reproducibility of studies and comparability of results across studies. 

%------------------------------------------------

\section*{Acknowledgments} 
\addcontentsline{toc}{section}{Acknowledgments} 

I would like to thank Dr Arthur White and Dr Jason Wyse for their support in the development of the ouladFormat R package. 

\bibliographystyle{apacite}
\bibliography{Bibliography}

%----------------------------------------------------------------------------------------

\end{document}